\documentclass[conference]{IEEEtran}
\IEEEoverridecommandlockouts
\usepackage{cite}
\usepackage{amsmath,amssymb,amsfonts}
\usepackage{algorithmic}
\usepackage{graphicx}
\usepackage{textcomp}
\usepackage{xcolor}

\usepackage[utf8]{inputenc}
\usepackage{kotex}
\usepackage{color}

\usepackage{comment}
\usepackage{caption}
\usepackage{subcaption}

\usepackage{tikz} 
\newcommand{\ballnumber}[1]{\tikz[baseline=(myanchor.base)] \node[circle,fill=.,inner sep=2pt] (myanchor) {\color{-.}\bfseries\footnotesize #1};} 
\usepackage{graphicx} 
\usepackage{color}
\definecolor{LightCyan}{rgb}{0.88,1,1}
\usepackage{colortbl, booktabs} 
\usepackage[T1]{fontenc}
\usepackage{listings}
\usepackage{xcolor} 
\usepackage{multicol}
\usepackage{multirow}
\usepackage{adjustbox}
\usepackage{changepage}
\usepackage{enumitem, kantlipsum}
\usepackage{tikz}
\usepackage[framemethod=default]{mdframed}

\usepackage{tabularx, tabulary, booktabs}
\usepackage{lipsum}
\usepackage[scaled=0.95]{beramono} 
\usepackage{hhline}
\usepackage{ctable}
\usepackage{makecell}
\usepackage[linesnumbered,ruled,vlined]{algorithm2e}
\usepackage{commath}
\usepackage{xspace}

\newcommand\nth{\textsuperscript{th}\xspace}
\usepackage{etoolbox}
\AtBeginEnvironment{algorithm}{\SetArgSty{textrm}}
\usepackage[hyphens]{url}
\usepackage[hidelinks]{hyperref}
\hypersetup{breaklinks=true}
\usepackage{float}

\def\BibTeX{{\rm B\kern-.05em{\sc i\kern-.025em b}\kern-.08em
    T\kern-.1667em\lower.7ex\hbox{E}\kern-.125emX}}

\begin{document}

\title{\texttt{LMStream}: When Distributed Micro-Batch Stream Processing Systems Meet GPU
}

\author{\IEEEauthorblockN{Suyeon Lee\IEEEauthorrefmark{2} and
Sungyong Park\IEEEauthorrefmark{1}}
\IEEEauthorblockA{Department of Computer Science and Engineering,
Sogang University\\
Seoul, South Korea\\
Email: \IEEEauthorrefmark{2}leesy0506@sogang.ac.kr, \IEEEauthorrefmark{1}parksy@sogang.ac.kr}}

\makeatletter
\patchcmd{\@maketitle}
  {\addvspace{0.5\baselineskip}\egroup}
  {\addvspace{-1.2\baselineskip}\egroup}
\makeatother

\maketitle

\begin{abstract}
\label{sec:abstract}
\textcolor{black}{
This paper presents \texttt{LMStream}, which ensures bounded latency while maximizing the throughput on the  GPU-enabled micro-batch streaming systems. The main ideas behind \texttt{LMStream}'s design can be summarized as two novel mechanisms: (1) dynamic batching and (2) dynamic operation-level query planning. By controlling the micro-batch size, \texttt{LMStream} significantly reduces the latency of individual dataset because it does not perform unconditional buffering only for improving GPU utilization. \textcolor{black}{\texttt{LMStream} bounds the latency to an optimal value according to the characteristics of the window operation used in the streaming application.} Dynamic mapping between a query to an execution device based on the data size and dynamic device preference improves both the throughput and latency as much as possible. In addition, \texttt{LMStream} proposes a low-overhead online cost model parameter optimization method without interrupting the real-time stream processing. 
We implemented \texttt{LMStream} on Apache Spark, which supports micro-batch stream processing. Compared to the previous throughput-oriented method, \texttt{LMStream} showed an average latency improvement up to a maximum of 70.7\%, while improving average throughput up to 1.74$\times$.}

\end{abstract}

\vspace{-0.05in}

\section{Introduction}\label{sec:introduction}

\footnote{Corresponding author: Sungyong Park*}
\textcolor{black}{
Big-data analysis has already become an essential part of various industries. With the explosive growth in data generated in real-time, stream processing, which provides fast and short-term results, has become a hot topic. In batch processing, the system processes large chunks of data at once with finite execution times. In contrast, a stream processing system continuously processes relatively small data without completing execution. In addition, stream data ingestion traffic is dynamic, which means that the size of the input data continuously changes at regular intervals~\cite{Q-Flink}\cite{Fine-Stream}. To efficiently handle these characteristics in a stream processing system, it  
should maximize the processing capacity of the entire system while maintaining the low latency of individual data. 
}

\textcolor{black}{
Recently, many stream processing systems have 
used GPUs 
to accelerate stream query processing either over a dedicated or an integrated CPU-GPU architecture~\cite{Fine-Stream}\cite{GPU-Spark}\cite{GPU-Flink}\cite{Saber}\cite{GPU-Online-Stream-Processing}\cite{Gasser}\cite{Toward-GPU-Accelerated-Streaming}.}
\textcolor{black}{
Especially, in a dedicated CPU-GPU architecture where memory is not shared and segregated per device, it has been known that there is data transition overhead (i.e., PCIe transfer time) between two different devices.  
Therefore, most previous works focused on a throughput-oriented approach 
that minimizes this overhead and increase GPU utilization~\cite{GPU-Spark}\cite{GPU-Flink}\cite{Saber}\cite{GPU-Online-Stream-Processing}\cite{Gasser}\cite{Toward-GPU-Accelerated-Streaming}.
These studies buffer real-time input datasets for a certain period and then transfer them to the GPU at once. As a result, they improve the throughput but significantly impair dataset-specific latency. We found that the unconditional static buffering in the existing micro-batch streaming model even induces an unlimited increase in the latency. Stream processing systems must address this issue because streaming services are latency-sensitive and often have specific requirements in this regard~\cite{Q-Flink}\cite{Workload-Aware-Streaming-State-Management}\cite{Cameo}.}



\textcolor{black}{
Since latency and throughput performance are trade-offs with each other, improving both simultaneously is challenging. To achieve this goal, we have established three design challenges. 
(1) We need a dynamic micro-batch controlling mechanism to ensure bounded latency because unconditional static buffering causes a significant increase in the latency.
(2) Since we need to reduce the total processing time to optimize latency and throughput simultaneously, 
an effective query planning mechanism is also required. 
(3) Additionally, when the stream processing application is first submitted, there is no prior information about the overall system performance based on the workload characteristics. Therefore, we need a mechanism to optimize parameters dependent on the resource configurations or workload types in an online manner. In this case, the running user application should not be interrupted.}

\textcolor{black}{
%
This paper presents \texttt{LMStream}, a GPU-enabled distributed micro-batch stream processing system that ensures bounded latency while maximizing the throughput. Our system design is composed of three parts to handle all the three challenges mentioned above effectively. First, \texttt{LMStream} performs dynamic batching to determine the optimal batch size to process at once per every micro-batch. \textcolor{black}{The system determines the optimal upper bound value of latency considering the characteristics of window operations essential for stream processing. Then, it adjusts the batch size so that the maximum latency of all micro-batches does not exceed this upper bound as much as possible.} Controlling the micro-batch size reduces maximum latency significantly, leading to a bounded tail latency. \textcolor{black}{Therefore, \texttt{LMStream} focuses on reducing the maximum latency to an appropriate value according to the system performance and application requirements.}
Second, \texttt{LMStream} performs an effective operation-level query planning to reduce the processing time as much as possible. It dynamically maps each query operation to an appropriate execution device (CPU or GPU). Dynamic device mapping is performed based on the pre-determined data size and device preference. Finally, \texttt{LMStream} conducts a low-overhead cost model parameter optimization in an online manner without interrupting real-time stream processing.}

\textcolor{black}{
We implemented \texttt{LMStream} using Apache Spark~\cite{Apache-Spark}\cite{Structured-Streaming}, one of the widely used distributed micro-batch stream processing systems. 
We integrated the cost models and algorithms proposed for \texttt{LMStream} into the original Spark and Spark-Rapids libraries~\cite{Spark-Rapids}. 
To demonstrate the efficiency of \texttt{LMStream}, we conducted experiments under various real-world workloads using different data ingestion traffics. Compared to the previous throughput-oriented method, \texttt{LMStream} showed an average latency improvement up to a maximum of 70.7\%, while improving average throughput up to 1.74$\times$.
}

\textcolor{black}{
To summarize, this paper makes the following specific contributions.
\textcolor{black}{\begin{itemize}
    \item{\texttt{LMStream} proposes a dynamic batching mechanism
    to ensure bounded maximum latency in the micro-batch streaming systems. 
    }
    \item{\texttt{LMStream} presents an effective operation-level query planning to improve both throughput and latency by utilizing each operation's dynamic device preference.  
    }
    \item{\texttt{LMStream} performs system parameter optimization online without interrupting real-time streaming applications.}
    \item{We implement a working system on Apache Spark and demonstrate its effectiveness through various real-world stream processing benchmarks.}
\end{itemize}}}

\section{Background and Motivation}\label{sec:backgroundAndMotivation}

\textcolor{black}{In this section, we first present the overview of distributed micro-batch stream processing system. Then, we explain how previous works utilize GPUs in the micro-batch model and discuss the limitations of those works. Lastly, we report several observations to motivate the main ideas of \texttt{LMStream}'s design.}

\subsection{Distributed Micro-batch Stream Processing Systems}\label{subsec:microBatchStreamPrcoessing}

\textcolor{black}{
The stream processing system ingests indefinitely generated data in real-time from the input stream and sends the computing result directly to the output stream. The input stream data may consist of various datasets with one or more files or row records. When new data are received, the system immediately performs simple inline analysis (e.g., moving average) and returns the necessary results to the user in a short time. Therefore, the stream processing mainly consists of a combination of several query operations.}

\textcolor{black}{
The stream processing systems follow either of the two execution models: event-driven model and micro-batch model. In the event-driven model, system pipelines query operations throughout the processing phase. The arriving data can be computed directly along the pipeline and thus each dataset has low latency. 
The execution unit is an individual dataset and this model achieves parallelism through pipelined computing.
In the micro-batch model, the focus is on the throughput, and not the latency. Therefore, when the new data arrive, they first enter the buffering phase and waits for the data to be collected for a certain period. When a trigger occurs, the collection of the dataset is transferred to the processing phase.
The execution unit in the micro-batch model is a single micro-batch, which is a collection of several datasets. The system first partitions the micro-batch and distribute partitioned data to CPU cores to execute the same query operations set on different data. Generally, the number of data partitions is the same as the number of CPU cores used per application.
The micro-batch model is superior in terms of throughput, however, the latency of the individual datasets is significantly compromised.}

\textcolor{black}{
The micro-batch model sets the query execution plan in advance before performing actual query processing. After the compiler analyzes the query and composes the operation directed-assigned-graph (DAG), the system determines an appropriate execution function per each operation. The method of how GPUs are utilized in the micro-batch streaming systems is simple. That is, the system just needs to specify that each operation uses the GPU-aware execution function. Using GPUs in the micro-batch model enables the entire system to achieve higher throughput. However, it exacerbates the latency problem of the micro-batch model by performing unconditional buffering. We have further described the details in Section~\ref{subsubsec:increasingLatency}.}

\subsection{Previous Works}\label{subsec:previousWorks}

\noindent{\textbf{Throughput-oriented Approach.}}\label{subsubsec:throughputOrientedApproach}
\textcolor{black}{
Previous studies using GPUs in the stream processing systems have commonly focused on the throughput. They postponed the query execution by buffering datasets to fully utilize the GPUs~\cite{GPU-Online-Stream-Processing}\cite{Gasser}\cite{Toward-GPU-Accelerated-Streaming}. Unconditional buffering causes the prolonged latency of individual datasets, which is fatal in streaming environments~\cite{Workload-Aware-Streaming-State-Management}\cite{Cameo}.
Another feature found in the throughput-oriented approaches is that systems offload query operations to the GPU as much as possible~\cite{GPU-Spark}\cite{GPU-Flink}\cite{Saber}\cite{GPU-Online-Stream-Processing} to avoid data transition overhead caused by the data movement between different devices.
However, it is well-known that distinct device preferences exist for each operation in the queries~\cite{Fine-Stream}\cite{Performance-Analysis-Of-ETL-Process}. Exploiting only a single device ignores those device preferences, thus failing to optimize the performance further.}

\vspace{0.1cm}
\noindent{\textbf{Approach Using Static Device Preference.}}\label{subsubsec:staticDevicePreference} \textcolor{black}{
Recently, FineStream~\cite{Fine-Stream} suggested an operation-level query planning over an integrated CPU-GPU architecture. However, applying the proposed idea to a dedicated architecture with inherent data transition overhead is hard. In addition, FineStream did not consider device preference that varies depending on the size of the data processed by the operation. Systems must reflect dynamic device preferences to optimize performance in streaming environments where the input data traffic fluctuates over time\cite{Q-Flink}\cite{Fine-Stream}.}

\subsection{Motivations}\label{subsec:ourObservations}

\noindent{\textbf{Faulty Pattern of Buffering: Endlessly Increasing Latency.}}\label{subsubsec:increasingLatency} \textcolor{black}{
Although the micro-batch model focuses on the throughput optimization, the essence of streaming is real-time processing; it needs to satisfy the minimum requirement for latency. However, in the case of unconditional buffering, the latency of each dataset exhibits an unlimited increase. To demonstrate this with a real-world application, we selected a single query of the Linear Road stream benchmark~\cite{Linear-Road-Bench} and ran it on the Apache Spark cluster that supports a micro-batch streaming model. We used constant data ingestion traffic, which indicates that the same-sized dataset is ingested every second. We have described the experimental setting and workload details in Section~\ref{subsec:experimentalSetup}.}

\begin{figure}[t]
    \centering\includegraphics[width=0.95\linewidth,keepaspectratio]{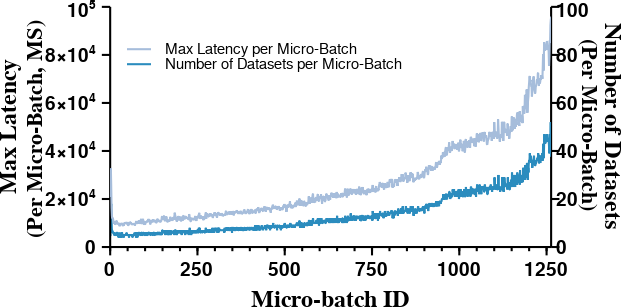}
    \vspace{-0.05in}
    \caption{Maximum of the latency values of each dataset per micro-batch, and the number of datasets per micro-batch.} 
    \vspace{-0.2in}
\label{fig:increasingLatency}
\end{figure}

\textcolor{black}{
As shown in Figure~\ref{fig:increasingLatency}, the maximum of the latency values in each dataset per micro-batch continues to increase as the micro-batch processing progresses. This phenomenon occurs because buffering takes place without considering the size of the data to be processed in a micro-batch. If the data size of the micro-batch exceeds the processing capacity, the time required for the processing phase increases, and the trigger interval becomes longer; correspondingly, the buffering phase time increases that much. In this case, since the amount of data to be processed in the subsequent micro-batch increases, the processing time also increases. Figure~\ref{fig:increasingLatency} shows that, similar to the increase in the latency, the number of datasets processed per micro-batch gradually increases. While this vicious cycle continues, the latency of individual datasets continues to increase. This also happens when using a GPU. Although utilizing a GPU can reduce the time in the processing phase, it cannot alleviate the increasing latency because, generally, a longer buffering time is required to fully utilize the GPU. Therefore, the latency value of each dataset should be bound in the micro-batch model by adjusting the size of the data processed in a single micro-batch.}

\begin{figure}[t]
    \centering\includegraphics[width=0.95\linewidth,keepaspectratio]{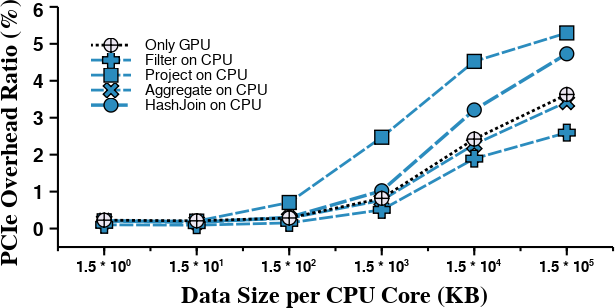}
    \vspace{-0.05in}
    \caption{PCIe overhead ratios for different batch data sizes and query operation types.}
    \vspace{-0.3in}
\label{fig:lowPCIeOverhead}
\end{figure}

\vspace{0.1cm}
\noindent{\textbf{Low PCIe Overhead for Small Data.}}\label{subsubsec:lowPCIeOverehadRatio} \textcolor{black}{
FineStream~\cite{Fine-Stream} advocated for an integrated CPU-GPU architecture because it can ignore the data transition overhead between different devices. However, we observed that the overhead is marginal if the data size is small. 
}

\textcolor{black}{Figure~\ref{fig:lowPCIeOverhead} shows the ratio of time required for the PCIe transmission to the total execution time, which is the leading factor of the data transition overhead. We used a synthetic select-project-join query for the following scenarios: (1) mapping all operations to the GPU, (2) mapping only a particular operation (i.e., filter or project) to the CPU and performing the rest on the GPU. We measured the PCIe transfer time using a GPU profiler called NVIDIA Nsight Systems. The graph shows that the PCIe overhead is minimal (i.e., less than 1\%) when processing small data, regardless of the query operation types. This result is noteworthy because it suggests that no force buffering of small streaming data is required to overcome the disadvantage of a dedicated CPU-GPU architecture. However, the overhead becomes significant once it starts processing data above a specific size. Therefore, the data transition overhead must be handled differently according to the data size to be processed.}

\section{Design}\label{sec:design}

\begin{figure*}[t]
    \centering\includegraphics[width=\linewidth,keepaspectratio]{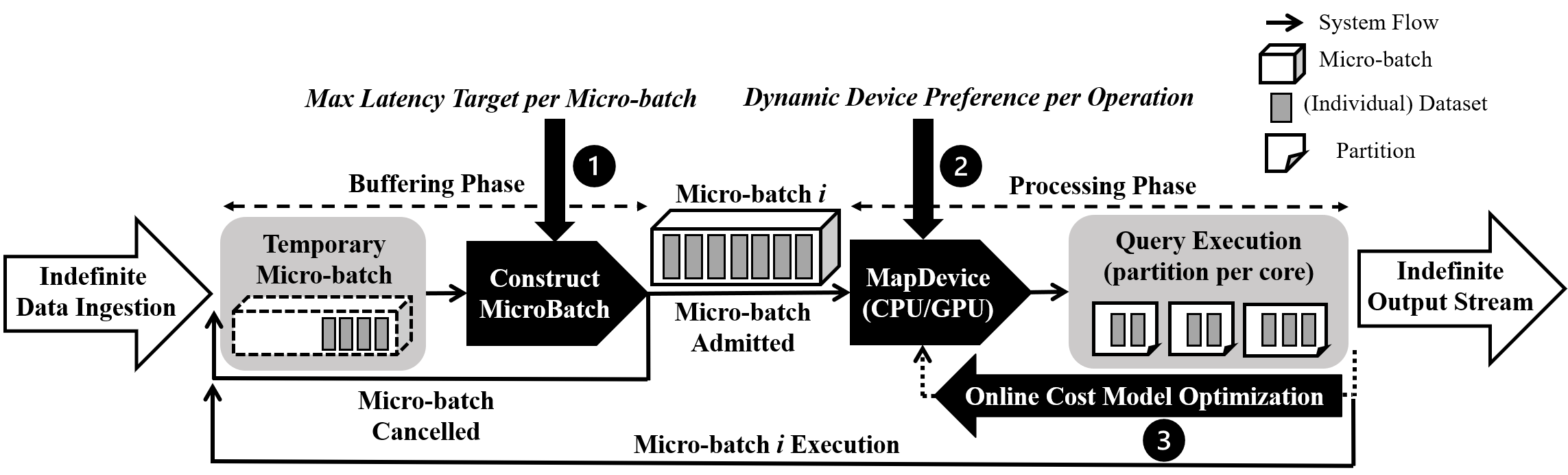}
    \vspace{-0.2in}
    \caption{\textcolor{black}{System overview of \texttt{LMStream}.}
    }
    \vspace{-0.05in}
\label{fig:overallSystems}
\end{figure*}
\setlength{\extrarowheight}{2pt}
\begin{table*}[t]
    \centering
    \caption{\textcolor{black}{Parameters used in cost models. All parameters are visible through the entire \texttt{LMStream} system.}}
    \vspace{-0.05in}
    \setlength\tabcolsep{2.5pt}
    \begin{tabular}{|c|c|c|}
        \hline
        \strut
        \textbf{Type} & \textbf{Notation} & \textbf{Description} \\
        \hline
        \multirow{3}{*}{Specified by User}
            & \multirow{1}{*}{$SlideTime$} & \makecell{Time value of window slide of a sliding window specified per query in user application. \\ If the value is 0, it means the application uses a tumbling window.} \\
            \cline{2-3}
            & \multirow{1}{*}{$NumCores$} & \makecell{A total number of CPU cores to process user application. \\ Usually, this value is as same as the number of data partitions.} \\
        \specialrule{.14em}{.1em}{.1em}
        \multirow{14}{*}{Defined in System}
            & \multirow{1}{*}{$NumDS_{i}$} & Total number of individual dataset in micro-batch $i$. \\
            \cline{2-3}
            & \multirow{1}{*}{$Part_{(i,j)}$} & Size of $j$\nth data partition in micro-batch $i$. \\
            \cline{2-3}
            & \multirow{1}{*}{$Buff_{(i,j)}$} & Time consumed in buffering phase for $j$\nth data partition in micro-batch $i$. \\
            \cline{2-3}
            & \multirow{1}{*}{$Proc_{i}$} & Time consumed in processing phase in micro-batch $i$. (All dataset in a single micro-batch has the same value.) \\
            \cline{2-3}
            & \multirow{1}{*}{$InfPT_{i}$} & \makecell{The value of inflection point used in micro-batch $i$ execution. \\ This value may vary for every micro-batch since it is optimized as stream processing continues.} \\
            \cline{2-3}
            & \multirow{1}{*}{$AvgThPut_{i}$} & Average throughput throughout entire streaming query obtained after execution of micro-batch $i$. \\
            \cline{2-3}
            & \multirow{1}{*}{$MaxLat_{i}$} & Max value among the latency of each dataset within micro-batch $i$.  \\
            \cline{2-3}
            & \multirow{1}{*}{$EstMaxLat_{i}$} & Estimated max value among the latency of each dataset within micro-batch $i$ (Estimation of $MaxLat_{i}$). \\
            \cline{2-3}
            & \multirow{1}{*}{$CPU_{(i,j,o)}$} & \makecell{CPU execution cost of operation $o$ to run $j$\nth data partition in micro-batch $i$. \\ When the operation prefers the CPU, its value is lower than corresponding GPU execution cost (i.e., $GPU_{(i,j,o)}$) .} \\
            \cline{2-3}
            & \multirow{1}{*}{$GPU_{(i,j,o)}$} & \makecell{GPU execution cost of operation $o$ to run $j$\nth data partition in micro-batch $i$. \\ When the operation prefers the GPU, its value is lower than corresponding CPU execution cost (i.e., $CPU_{(i,j,o)}$).} \\
            \cline{2-3}
            & \multirow{1}{*}{$Trans_{(i,j,o)}$} & \makecell{Data transition cost of operation $o$ to run $j$\nth data partition in micro-batch $i$. \\ This value exists when the current operation $o$ runs on a different device than the previous operation $o-1$.} \\
            \cline{2-3}
            \hline
    \end{tabular}
    \vspace{-0.2in}
\label{tab:costModelParams}
\end{table*}

\subsection{Overview of \texttt{LMStream}}\label{subsec:overallSystems}

\textcolor{black}{
A micro-batch execution in stream processing system consists of two phases: buffering phase and processing phase. The buffering phase waits for new data for a constant period. When the trigger occurs, the ingested data forms a single micro-batch, and the system moves the micro-batch to the processing phase to perform query executions. The execution unit is not a single dataset but a single micro-batch that contains a certain number of datasets. The trigger is a time value provided by the user, which indicates the interval of processing phase. The system idles actual query execution for the trigger time and composes a micro-batch, regardless of how early the previous execution ends.}



\textcolor{black}{
In \texttt{LMStream}, we deprecate the trigger concept since its static buffering phase time deteriorates the overall system performance as shown in Section~\ref{subsec:ourObservations}. Therefore, \texttt{LMStream} works quite differently from the original micro-batch model. Figure~\ref{fig:overallSystems} shows the overall system architecture of \texttt{LMStream}. The system consists of three core parts. First, we designed a micro-batch admission controller called \texttt{ConstructMicroBatch} (\ballnumber{1}) to bound the latency to the optimal value. Second, we designed an operation-level query planner called \texttt{MapDevice} (\ballnumber{2}) to perform with the optimal processing phase time. Lastly, \texttt{LMStream} optimizes several cost model parameters online, considering different system configurations and workload types (\ballnumber{3}).}


\textcolor{black}{
\ballnumber{1} Instead of using the trigger concept, \texttt{LMStream} runs the \texttt{ConstructMicroBatch} module such that the system can determine a suitable size of a micro-batch. \textcolor{black}{The module uses cost models considering streaming window requirements set by users in the application.} By adopting \texttt{LMStream}'s micro-batch controlling mechanism, the system can dynamically adjust the buffering phase time and the micro-batch size. If the \texttt{ConstructMicroBatch} module admits the micro-batch, the system passes the micro-batch to a \texttt{MapDevice} module to determine the operation-level query execution plans. On the other hand, if the \texttt{ConstructMicroBatch} module cancels the micro-batch, the involved datasets step back and wait for a re-request to the \texttt{ConstructMicroBatch} module. In our implementation, whenever there is no valid micro-batch, the \texttt{ConstructMicroBatch} module is called every ten milliseconds to poll data from the input data traffic and request new micro-batch admission.}

\textcolor{black}{
\ballnumber{2} In the following query planning process, \texttt{LMStream} runs a \texttt{MapDevice} module that creates an operation-level CPU-GPU execution plan. Device selection reflects the device preferences of each operation. These device preferences depend on the micro-batch data size. Since the data size of a single micro-batch varies among the fluctuating input data traffics, the device mapping results can be different for every micro-batch execution.} 

\textcolor{black}{
\ballnumber{3} Initially, \texttt{LMStream} initializes the parameters used in cost models with the static values based on the pre-experimental results. As the stream processing continues, \texttt{LMStream} optimizes the parameter values using two types of information: performance history information and target performance information. The parameter optimization process is an asynchronous background process separate from the main micro-batch execution flow. This design can minimize the overhead imposed on the overall streaming system.}

\subsection{Problem Definition}\label{subsec:problemDefinition}

\textcolor{black}{
The parameters used for the cost models and algorithms are summarized in Table~\ref{tab:costModelParams}.
Unlike previous throughput-oriented methods, the whole system goal of \texttt{LMStream} is to maximize the average throughput while keeping the latency bound to an appropriate value. 
Thus, our goal can be represented as below in Equation~\ref{eq:objective},~\ref{eq:constraint1}, and~\ref{eq:constraint2}.
%
\begin{alignat}{3}
&\!\max_{i}        &\quad& AvgThPut_{i}\label{eq:objective}\\
&\text{s.t.} &\quad&MaxLat_{i}<SlideTime&\quad (SlideTime>0)\label{eq:constraint1}\\
&                  &\quad&MaxLat_{i}\leq \frac{\sum_{k=0}^{i-1}MaxLat_{k}}{i-1}&\quad (SlideTime=0)\label{eq:constraint2}
\end{alignat}
%
To achieve this goal, \texttt{LMStream} updates $AvgThPut_{i}$ and $MaxLat_{i}$ after every execution of micro-batch $i$. Each metrics can be defined as Equation~\ref{eq:averageThroughput} and Equation~\ref{eq:maxLatency} in a distributed micro-batch stream processing model.
%
\begin{equation}
\begin{gathered}
    AvgThPut_{i} = \frac{\sum_{k=0}^{i}\sum_{j=0}^{NumCores}Part_{(k,j)}}{\sum_{k=0}^{i}Proc_{k}}
\end{gathered}
\label{eq:averageThroughput}
\end{equation}
\begin{equation}
\begin{gathered}
    MaxLat_{i} = \max_{j \in NumDS_{i}}(Buff_{(i,j)})+Proc_{i}
\end{gathered}
\label{eq:maxLatency}
\vspace{-0.15in}
\end{equation}}

\textcolor{black}{
The throughput and the latency are in a trade-off relationship. To maximize the throughput, system should maximize the data size of a single micro-batch (i.e., $Part_{(i,j)}$ in Equation~\ref{eq:averageThroughput}). However, this leads to longer buffering phase time (i.e., $Buff_{(i,j)}$ in Equation~\ref{eq:maxLatency}) and increases the latency. 
To optimize the two factors simultaneously, the overall mechanism of \texttt{LMStream} progresses two steps. (1) It determines the optimal data size of a single micro-batch by being aware of window types (referring to $SlideTime$). (2) It maps each query operation to an appropriate device (CPU or GPU) based on the micro-batch data size so that system can perform with minimum processing time. Module \texttt{ConstructMicroBatch} is for step (1), and module \texttt{MapDevices} is for step (2).}

\subsection{Construct Micro-batch}\label{subsec:constructMicroBatch}

\textcolor{black}{
This section describes the cost models and algorithm in \texttt{ConstructMicroBatch} module to determine the optimal size of micro-batch per execution. The algorithm detail is shown in \textbf{Algorithm 1}. By default, \texttt{ConstructMicroBatch} polls unprocessed input datasets every ten milliseconds. If the new data has not arrived, the module returns an empty micro-batch immediately to the main progress to continue polling.
If the new data exists, they ($newFiles$) are sorted by creation time and compose a temporary micro-batch ($tmpMicroBatch$). Temporary micro-batches include any pending data from previously canceled micro-batches ($bufferedFiles$). Temporary micro-batch is yet invalid, and \texttt{ConstructMicroBatch} starts deciding whether to validate it. If it judges that the system needs to process the corresponding data immediately, it admits the temporary micro-batch and passes it to the processing phase. Otherwise, it cancels the temporary micro-batch and continues polling to collect more data. In this case, a buffering phase time keeps increasing.}

\SetKwInput{KwInput}{Input}
\SetKwInput{KwOutput}{Output}
\SetKw{Break}{Break}
\SetKw{True}{True}
\SetKw{False}{False}
\let\emptyset\varnothing
\newcommand\mycommfont[1]{\footnotesize\ttfamily\textcolor{blue}{#1}}
\SetCommentSty{mycommfont}
\let\emptyset\varnothing
\begin{algorithm}[t]
\DontPrintSemicolon
\SetAlgoLined
\caption{Control admission of micro-batch to determine optimal data size per execution.}
    \KwResult{(\textbf{<Boolean>} Admission request result, \textbf{<Micro-batch>}  Admitted micro-batch, \textbf{<Micro-batch>} Canceled micro-batch)}
    \SetKwFunction{FConstructMicroBatch}{ConstructMicroBatch}
    \SetKwProg{Fn}{Def}{:}{}
    \Fn{\FConstructMicroBatch{}}{
        \uIf{there is no new data}{
            \tcp*[h]{do polling}\;
            \KwRet (\False, $\emptyset$, $\emptyset$)\;
        }
        
        Get all new data in the source path as $newFiles$\;
        Sort $newFiles$ by creation time\;
        Load $bufferedFiles$\;
        $tmpMicroBatch$ = $bufferedFiles$ $\cup$ $newFiles$
        
        
        
        \uIf(\tcp*[h]{Sliding Window}){$SlideTime$ > 0}{
            \uIf{$EstMaxLat_{i} \geq SlideTime$}{
                \tcp*[h]{process immediately}\;
                $bufferedFiles$ = $\emptyset$\;
                \KwRet (\True, $tmpMicroBatch$, $\emptyset$)\;
            }
        }
        \uElseIf( \tcp*[h]{Tumbling Window}){$SlideTime$ == 0}{
            \uIf{$EstMaxLat_{i} \geq \frac{\sum_{k=0}^{i-1}MaxLat_{k}}{i-1}$}{
                \tcp*[h]{process immediately}\;
                $bufferedFiles$ = $\emptyset$\;
                \KwRet (\True, $tmpMicroBatch$, $\emptyset$)\;
            }
        }
        
        \tcp*[h]{do buffering}\;
        $bufferedFiles$ = $tmpMicroBatch$\;
        \KwRet (\False, $\emptyset$, $tmpMicroBatch$)\;
    }
\setlength{\textfloatsep}{0pt}
\end{algorithm}

\begin{figure}[t]
\vspace{-0.15in}
    \centering\includegraphics[width=\linewidth,keepaspectratio]{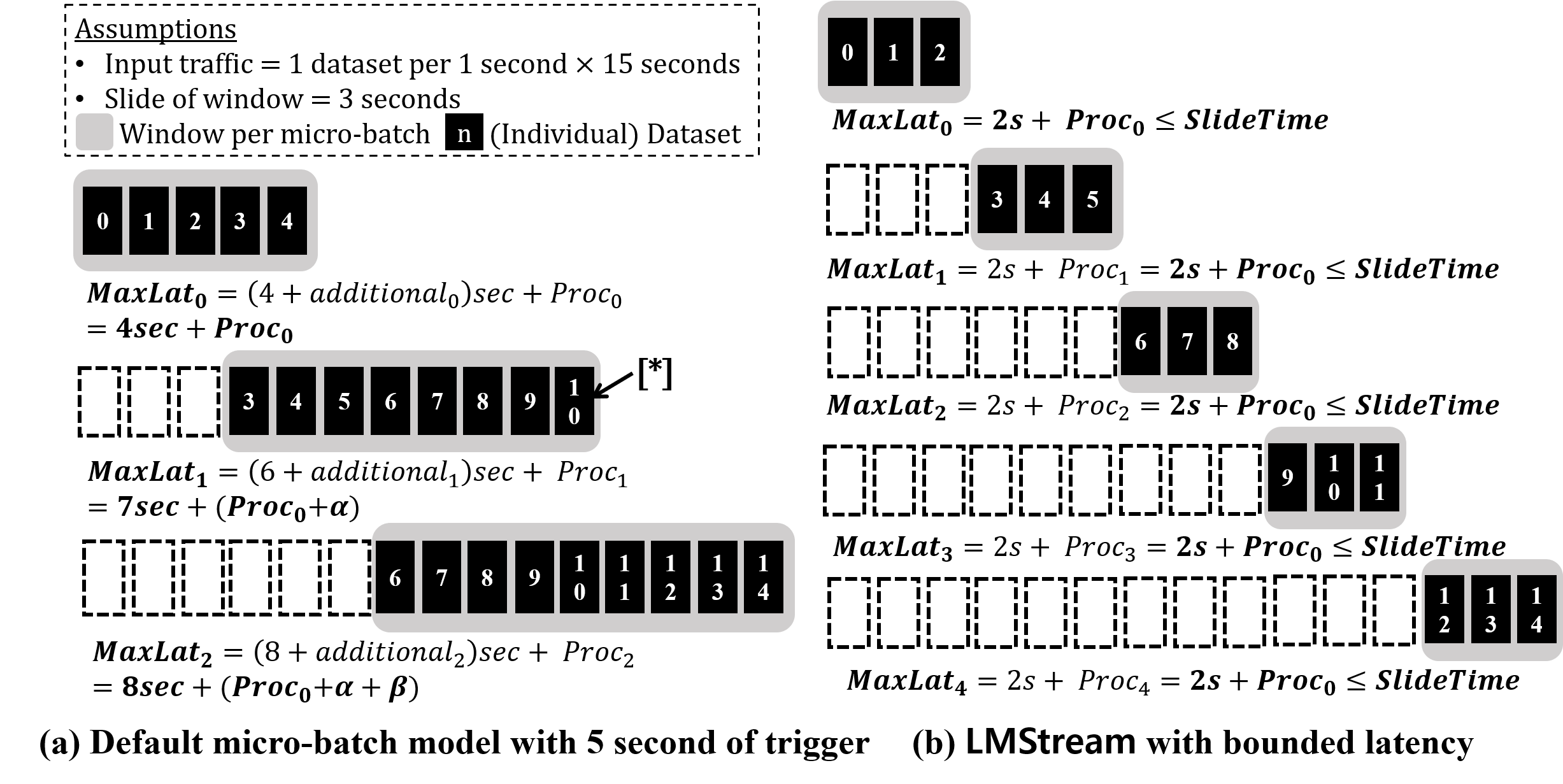}
    \vspace{-0.2in}
    \caption{\textcolor{black}{Example scenario to demonstrate the need that the maximum latency of a micro-batch should be bound to the slide time of the window. [*] is for presenting a situation where $additional_{i}$ occurs. In this figure, $proc_{0}$ is 1 second longer than the trigger 5 seconds, then the input traffic has more than five datasets.}}
    \vspace{-0.3in}
\label{fig:slideAwareBatching}
\end{figure}

\vspace{0.1cm}
\noindent{\textbf{Window slide time as deadline.}}\label{subsubsec:slideSizeAsDeadline} \textcolor{black}{
In this paper, we set the optimal upper bound of maximum latency as the slide time of the window ($slideTime$). If the slide time is greater than 0, it means that the streaming application performs a sliding window operation. In this case, if the data amount per micro-batch is not adjusted properly, its size will gradually increase, and latency might be greatly impaired. Figure~\ref{fig:slideAwareBatching} shows the scenario through a simple example. Suppose that a single dataset is continuously coming in per second, and the slide time of the window is 3 seconds. Figure~\ref{fig:slideAwareBatching}(a) is an default micro-batch model with a 5-second trigger. In this case, since the number of the performed dataset according to the slide time is smaller than the amount of data added in a new micro-batch, the data to be processed per micro-batch gradually increases as streaming progresses. As the data increases, the time required for the processing phase ($Proc_{i}$) also increases, so additional data ($additional_{i}$ in Figure~\ref{fig:slideAwareBatching}) might be accumulated during that time, increasing the data to be processed further. As a result, the maximum latency ($maxLat_{i}$) per micro-batch increases rapidly. Figure~\ref{fig:slideAwareBatching} only shows the case where the input data rate is constant, but when the input data traffic becomes dynamic, this problem can become more severe. On the other hand, if the maximum latency of micro-batch is bound to the slide time of the window regardless of input data traffic, the latency value can be maintained as shown in Figure~\ref{fig:slideAwareBatching}(b). Therefore, \texttt{LMStream} maintains the maximum latency of each micro-batch to be almost similar but less than the corresponding slide time when the slide time is greater than 0 (Equation~\ref{eq:constraint1}). If the slide time is 0, it means that the streaming application performs a tumbling window operation. In this case, \texttt{LMStream} maintains the maximum latency of each micro-batch by converging to the average of its past results (Equation~\ref{eq:constraint2}).}

\textcolor{black}{
\texttt{ConstructMicroBatch} module estimates the maximum latency of a micro-batch and compares it with the latency value that the system targets to maintain. Then it decides whether to execute the micro-batch immediately or wait a little longer. The cost model to predict the maximum latency of micro-batch is as below Equation~\ref{eq:estimatedLatency}.
\begin{equation}
    EstMaxLat_{i} = \max_{j \in NumDS_{i}}(Buff_{(i,j)}) + \frac{\sum_{j=0}^{NumDS_{i}}Part_{(i,j)}}{AvgThPut_{i-1}} \\
\label{eq:estimatedLatency}
\end{equation}
If $EstMaxLat_{i}$ is larger or equal to $slideTime$ or its average value, \texttt{ConstructMicroBatch} admits the temporary micro-batch and starts processing phase immediately to satisfy the constraints of Equation~\ref{eq:constraint1} and Equation~\ref{eq:constraint2}. Otherwise, the temporary micro-batch will be canceled, and the module stores the data to be used in the next micro-batch admission judge round.}

\subsection{Map Device}\label{subsec:mapDevices}

\textcolor{black}{This section describes the cost models and algorithm to map an optimal device (CPU or GPU) dynamically per operation according to the determined micro-batch data size. Before presenting the details of the mechanism, we first show the necessity of dynamic device preference by data size and a key concept named \textit{inflection point}.}

\begin{figure}[t]
    \centering\includegraphics[width=0.95\linewidth,keepaspectratio]{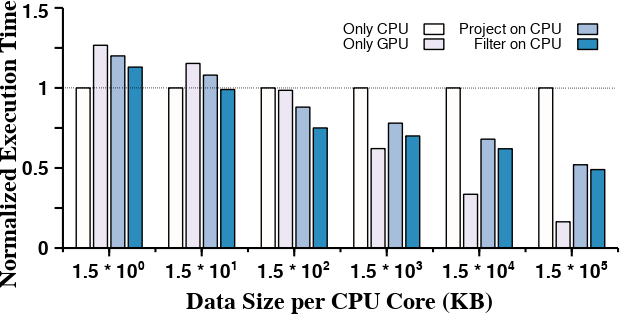}
    \vspace{-0.05in}
    \caption{Normalized execution times for different batch data sizes and query operation types.}
    \vspace{-0.2in}
\label{fig:necessityOfDynamicDevicePreference}
\end{figure}

\vspace{0.1cm}
\noindent{\textbf{Necessity of Dynamic Device Preference.}}\label{subsubsec:necessityDynamicDevicePreference} \textcolor{black}{
Figure~\ref{fig:necessityOfDynamicDevicePreference} shows the normalized execution times of a same synthetic query used in Section~\ref{subsec:ourObservations} in case of the following scenarios: (1) mapping all operations to the CPU, (2) mapping all operations to the GPU, (3) mapping only a particular operation (i.e., filter or project) to the CPU and performing all others on the GPU. The normalization was performed based on the execution time of the scenario (1). As shown in the bar graph, some operations prefer the CPU if the batch data size is smaller than a threshold value (i.e., around 15 KB to 150 KB). However, if it exceeds the threshold, all operations start to prefer the GPU. When the data size is too small (i.e., less than 15KB), it is best to use only the CPU. If the data size reaches a specific size (i.e., 150 KB), the performance is better when the CPU and GPU are mixed rather than using any single device. As the data size increases, CPU affinity drops sharply. Thus, it is appropriate to perform all operations on the GPU. Experiments on more diverse operations can be found in our previous work~\cite{Performance-Analysis-Of-ETL-Process}, which shows that dynamic device preference is essential. In particular, it is more critical in a streaming environment where the input data traffic varies continuously.}

\vspace{0.1cm}
\noindent{\textbf{Inflection Point.}}\label{subsubsec:inflectionPointNotes} \textcolor{black}{
Mapping each operator differently to a CPU or GPU creates a performance branch according to specific data size. What's more noteworthy is that the point where data transition overhead begins to surge is as same as the performance branch point. In this paper, the data size that initiates those branch points is called \textit{inflection point}. We use the idea of specifying dynamic device preferences according to input data size contingent on inflection points. \texttt{LMStream} uses its initial value as 150 KB and optimizes gradually during stream processing running. We describe the detail of the online optimization process in Section~\ref{subsec:parameterOptimizations}.}

\setlength{\extrarowheight}{2pt}
\begin{table}[t]
    \centering
    \caption{\textcolor{black}{Initial preference~\cite{Performance-Analysis-Of-ETL-Process} and base cost of each operation.}}
    \vspace{-0.05in}
    \begin{tabular}{|c|c|c|}
        \hline
        \strut
        \textbf{Query Operation} & \textbf{Initial Preference} & \textbf{Base Cost} \\
        \hline
        \multirow{1}{*}{Aggregation (Hash)} & {CPU} & {1.0} \\
        \hline
        \multirow{1}{*}{Filtering} & {CPU} & {1.0} \\
        \hline
        \multirow{1}{*}{Shuffling} & {CPU} & {1.0} \\
        \hline
        \multirow{1}{*}{Projection} & {Neutral} & {0.9} \\
        \hline
        \multirow{1}{*}{Join (Hash)} & {Neutral} & {0.9} \\
        \hline
        \multirow{1}{*}{Expand} & {Neutral} & {0.9} \\
        \hline
        \multirow{1}{*}{Scan (CSV File)} & {GPU} & {0.8} \\
        \hline
        \multirow{1}{*}{Sorting} & {GPU} & {0.8} \\
        \hline
    \end{tabular}
    \vspace{+0.1in}
\label{tab:staticDevicePreference}
\end{table}

\vspace{0.1cm}
\noindent{\textbf{Dynamic Device Preference by Data Size.}}\label{subsubsec:dynamicDevicePreference} \textcolor{black}{
To reduce processing phase time as much as possible, \texttt{MapDevice} module maps an optimal device (CPU or GPU) per operation according to the determined micro-batch data size. The algorithm detail is shown in \textbf{Algorithm 2}. First, the micro-batch streaming system expresses query operations as DAG according to the execution order and data dependencies. Then, \texttt{MapDevice} maps each operation to the execution function of the optimal device. Initially, all operations are mapped to the GPU. If a specific operation has the execution cost of CPU function smaller than that of GPU function, \texttt{MapDevice} changes the operation's device mapping to the CPU.}

\SetKwInput{KwInput}{Input}
\SetKwInput{KwOutput}{Output}
\SetKw{Break}{Break}
\SetKw{True}{True}
\SetKw{False}{False}
\let\emptyset\varnothing
\begin{algorithm}[t]
\DontPrintSemicolon
\SetAlgoLined
\caption{Map suitable device per operation.}
    \KwResult{\textbf{<Query Plan>} Query execution plan with optimal device mappings}
    \SetKwFunction{FMapDevice}{MapDevice}
    \SetKwProg{Fn}{Def}{:}{}
    \Fn{\FMapDevice{}}{
        Get DAG of query operations as $opDAG$\;
        Initially, map every operation in $opDAG$ to the GPU and set as $queryPlan$\;
        \ForEach{$o \in$ \text{traverse($queryPlan$)}}{
            Calculate $CPU_{(i,j,o)}$ and $GPU_{(i,j,o)}$\;
            \uIf{($o$ is first operation) or ($o$ is last operation) or ($o-1$ is running on the CPU)}{
                $GPU_{(i,j,o)}$ += $Trans_{(i,j,o)}$\;
            }
            \uElse{
                $CPU_{(i,j,o)}$ += $Trans_{(i,j,o)}$\;
            }
            \uIf{$GPU_{(i,j,o)}$ > $CPU_{(i,j,o)}$}{
                Map $o$ to run on the CPU\;
            }
        }
        \KwRet $queryPlan$\;
    }
\end{algorithm}
\setlength{\textfloatsep}{0pt}

\vspace{-0.05in}
\textcolor{black}{Searching sequentially from the child node of DAG, it sets the base cost of each operations ($baseCost_{o}$ in Equation~\ref{eq:cpuPreference} and Equation~\ref{eq:gpuPreference}) and calculate execution cost of each device's function. According to our prior work~\cite{Performance-Analysis-Of-ETL-Process} to reveal general device preference per operation, we set base costs as Table~\ref{tab:staticDevicePreference}. In the table, initial preference is device preference of each operation when data size is as similar as \textit{inflection point}. When \texttt{MapDevice} applies dynamic device preference according to the data size, the size of data processed per one CPU core is the size of the data partition, not the size of a single micro-batch. As shown in Figure~\ref{fig:overallSystems}, the micro-batch streaming system first partitions a single micro-batch as many as the number of CPU cores. Data partitions are distributed and processed parallelly. Therefore, each device function's execution cost can be expressed as Equation~\ref{eq:cpuPreference} and Equation~\ref{eq:gpuPreference}.
\vspace{-0.05in}
\begin{equation}
    CPU_{(i,j,o)} = baseCost_{o} \times (\frac{Part_{(i,j)}}{InfPT_{i}})
\label{eq:cpuPreference}
\end{equation}
\vspace{-0.1in}
\begin{equation}
    GPU_{(i,j,o)} = baseCost_{o} \times (\frac{InfPT_{i}}{Part_{(i,j)}})
\label{eq:gpuPreference}
\vspace{-0.05in}
\end{equation}
If the size of the data partition is bigger than $InfPT_{i}$, the GPU's execution cost will be lower. On the contrary, if the size of the data partition is smaller than $InfPT_{i}$, the CPU's execution cost will be lower. Each operation will eventually select the device with a lower execution cost.}

\textcolor{black}{
While determining the current operation's device selection, it is crucial to consider the previous operation's device selection. Data transition overhead occurs when a previous operation and a current operation are running on different devices. As mentioned, this overhead begins to surge when the data size becomes bigger than the \textit{inflection point}. Therefore, we define the data transition cost ($Trans_{(i,j,o)}$) as Equation~\ref{eq:transitionCost} with the same mechanism as when determining dynamic device preference.
\begin{equation}
\begin{gathered}
    Trans_{(i,j,o)} = baseTransCost \times (\frac{Part_{(i,j)}}{InfPT_{i}}) \\
\end{gathered}
\label{eq:transitionCost}
\vspace{-0.05in}
\end{equation}
We set initial baseTransCost as 0.1. If the operation is the leaf or root of the DAG, the system needs to fetch/load data from/to the CPU. Therefore, \texttt{MapDevice} adds $Trans_{(i,j,o)}$ to $GPU_{(i,j,o)}$. Also, if previous operation is mapped to the CPU, it adds $Trans_{(i,j,o)}$ to $GPU_{(i,j,o)}$. Otherwise, it adds $Trans_{(i,j,o)}$ to $CPU_{(i,j,o)}$. Finally, if the value of $GPU_{(i,j,o)}$ is greater than $CPU_{(i,j,o)}$, meaning that the execution cost of GPU is greater than that of the CPU, then \texttt{MapDevice} maps the operation to run on the CPU.}

\subsection{Parameter Optimization}\label{subsec:parameterOptimizations}

\textcolor{black}{
\textit{Inflection point} is initialized with the values derived from the experiments that are dependent on resource configurations or workload types. Since the CPU/GPU device mapping per operation mainly varies based on \textit{inflection point}, \texttt{LMStream} should re-assign an optimal value within its system without blocking streaming applications.
To this end, \texttt{LMStream} internally performs regression using the following Equation~\ref{eq:regressionModel}.
\begin{equation}
\begin{gathered}
    InflectionPoint = \beta_0 + \beta_1\, Throughput + \beta_2\, Latency
\end{gathered}
\label{eq:regressionModel}
\end{equation}
\texttt{LMStream} performs regression at the end of every micro-batch execution. Therefore, \textit{inflection point} might be updated every micro-batch ($InfPT_{i}$). $\beta_0 $, $\beta_1$, and $\beta_2$ are coefficients which are determined every end of micro-batch execution accordingly. We use the simplest yet powerful model with appropriate input data and target values to avoid interrupting real-time streaming applications. Input data are histories of average throughput per micro-batch and maximum dataset-specific latency per micro-batch. For this purpose, \texttt{LMStream} tracks the information of past micro-batches.
In addition, \texttt{LMStream} dynamically sets target throughput and target latency according to the current system situation. It sets target throughput as the maximum value among previous data. It sets target latency to the window slide time (following Equation~\ref{eq:constraint1}), or sets it to the average value among previous data (following Equation~\ref{eq:constraint2}). These values become test inputs of regression so that the system can infer an optimal inflection point for the subsequent micro-batch execution.}

\setlength{\extrarowheight}{2pt}
\begin{table*}
    \centering
    \setlength\tabcolsep{3.5pt}
    \caption{Query details of real-world streaming workloads used in our experiments.}
    \vspace{-0.05in}
    \begin{tabular}{|c|c|c|c|}
        \hline
        \strut
        \textbf{Benchmark} & \textbf{Notation} & \textbf{\makecell{Window \\ Type}} & \textbf{Query Details} \\
        \hline
        \multirow{4}{*}{\makecell{Linear \\ Road \\ ~\cite{Linear-Road-Bench}}}
            & \multirow{1}{*}{$LR1S$} & Sliding & \multirow{2}{*}{\makecell{SELECT L.timestamp, L.vehicle, L.speed, L.highway, L.lane, L.direction, L.segment \\ FROM SegSpeedStr [range 30 (slide 5)] as A, SegSpeedStr as L WHERE (A.vehicle == L.vehicle)}} \\
            \cline{2-3}
            & \multirow{1}{*}{$LR1T$} & Tumbling & \\
            \cline{2-4}
            & \multirow{1}{*}{$LR2S$} & Sliding & \makecell{SELECT timestamp, highway, direction, segment, AVG(speed) as avgSpeed \\ FROM SegSpeedStr [range 30 slide 10] GROUPBY (highway, direction, segment) HAVING (avgSpeed < 40.0)} \\
        \specialrule{.14em}{.1em}{.1em}
        \multirow{3}{*}{\makecell{Cluster \\ Monitoring \\ ~\cite{Google-Cluster-Monitoring-Bench}}}
            & \multirow{1}{*}{$CM1S$} & Sliding & \multirow{2}{*}{\makecell{SELECT timestamp, category, SUM(cpu) as totalCpu \\ FROM TaskEvents [range 60 (slide 10)] GROUPBY category ORDERBY SUM(cpu)}} \\
            \cline{2-3}
            & \multirow{1}{*}{$CM1T$} & Tumbing & \\
            \cline{2-4}
            & \multirow{1}{*}{$CM2S$} & Sliding & \makecell{SELECT jobId, AVG(cpu) as avgCpu \\ FROM TaskEvents [range 60 slide 5] WHERE (eventType == 1) GROUPBY jobId} \\
            \hline
    \end{tabular}
    \vspace{-0.2in}
\label{tab:queryDetails}
\end{table*}

\textcolor{black}{
\texttt{LMStream} handles the optimization process asynchronously. After the query execution completion, stream processing needs to run additional tasks such as check-pointing and state flushing. Since the optimization process is performed during this period, and because the results only need to be returned before the next processing phase (as shown in Figure~\ref{fig:overallSystems}), the optimization process rarely blocks real-time streaming applications. Its overhead does not significantly affect the overall performance of the stream processing system.
However, assuming that stream processing continues indefinitely, the size of the historical data used for regression training will increase as much as that amount. 
Further studies can introduce various policies, such as periodic optimization or using only the latest $N$ data for the regression training.}

\section{Implementation}\label{sec:implementation}

\textcolor{black}{
We have implemented \texttt{LMStream} on Apache Spark~\cite{Apache-Spark} that supports micro-batch type stream processing. In particular, Spark SQL~\cite{Spark-SQL}, one of the native core modules, supports SQL-based optimized query processing\cite{Structured-Streaming}. Spark creates a single micro-batch of ingested input data for every trigger and creates a query plan. Query planning has two steps: (1) a logical planning step that creates a DAG of query operations, (2) a physical planning step that maps a specific function (algorithm) per operation. If the system uses GPUs, NVIDIA's Spark-Rapids library~\cite{Spark-Rapids} maps the operation with a GPU function in the physical planning stage. The Spark-Rapids library has implemented the GPU functions for almost all operations and manages the operation mapping process in connection with Spark. The default configuration designates the trigger as a static time interval (immutable during runtime) and maps possible GPU functions unconditionally to all query operations. That is, the primary mechanism of the original Spark is as same as the previous throughput-oriented method.}

\textcolor{black}{
To implement \texttt{LMStream}, we first set the trigger value as zero to ignore the trigger concept and implemented the \texttt{ConstructMicroBatch} module into the original Spark. 
Next, we implemented the \texttt{MapDevice} module in the Spark-Rapids library. Finally, we made the online cost model optimization process triggered in Spark. The simple regression process runs asynchronously by using Scala's \textit{ProcessBuilder} and \textit{Future}. We mainly used Scala to implement overall \texttt{LMStream} and wrote only the regression process in Python. The source codes are available at \textit{https://github.com/suyeon0506/LMStream}.}
\section{Evaluation}\label{sec:evaluation}

\subsection{Experimental Setup}\label{subsec:experimentalSetup}

\textcolor{black}{
To show the \texttt{LMStream}'s efficiency, we first evaluated the overall performance compared to the \texttt{Baseline}. Then we performed a microanalysis of each of the three core mechanisms used in our design. \texttt{Baseline} is the default Spark with Spark-Rapids using the existing throughput-oriented method as described in Section~\ref{sec:implementation}. It has a trigger time of 10 seconds, which value is the same or slightly larger than the window slide size of the workload used in our experiments. We repeated three times for all experiments and used their average value as the final result value.
}

\vspace{0.1cm}
\noindent{\textbf{Configurations.}}\label{subsubsec:expConfigurations} \textcolor{black}{
For the experiment, we configured a Spark cluster consisting of one master node and two worker nodes. We placed two executors per worker node. Each executor uses 12 CPU cores, 24 GB host memory, 1 GPU, and 8 GB GPU device memory as resources. The types of CPU and GPU are the Intel Xeon Silver 4210 and the NVIDIA GeForce RTX 2080 Ti, respectively.}

\vspace{0.1cm}
\noindent{\textbf{Workloads and Stream Traffic Types.}}\label{subsubsec:expWorkloads} \textcolor{black}{
The workloads used in the experiments are summarized in Table~\ref{tab:queryDetails}. The Linear Road benchmark~\cite{Linear-Road-Bench} and Cluster Monitoring benchmark~\cite{Google-Cluster-Monitoring-Bench} are real-world streaming benchmarks that perform various query operations, including filter, project, shuffle aggregate, join, etc. By modifying the query used in previous studies~\cite{Fine-Stream}\cite{Saber}, we varied the window slide size and added frequently used query operations such as sort. We used the following types of input data traffic for our experiments. Both traffic transfers enough data, fully loading the computing capacity of the cluster.
\begin{itemize}
    \item \textbf{Constant Traffic}: Every second, 1000 data rows are ingested as one dataset. In the case of a linear road workload, the size of one dataset input every second is about 60$\sim$70 KB. In the case of cluster monitoring workload, the size of one dataset input every second is about 150$\sim$200 KB. We used the traffic for fair comparison when evaluating overall performance.
    \item \textbf{Random Traffic}: In a realistic streaming environment, input data traffic constantly fluctuates. We figured this traffic to show that dynamic batching and query planning of \texttt{LMStream} works practically in a real-world environment. Every second, a random number of data rows are ingested and form a dataset. These random numbers are generated every second, achieving a normal distribution of 1000 as an average point.
\end{itemize}}

\subsection{Overall Performance}\label{subsec:overallPerformance}

\textcolor{black}{
This section compares the primary performance of \texttt{LMStream} with those of \texttt{Baseline}. After running long enough until the performance of \texttt{LMStream} converges to a specific value for each query execution, we measured the average dataset-specific end-to-end latency and the average throughput of each system. For a fair comparison, we used constant traffic.}

\begin{figure}[t]
    \centering\includegraphics[width=\linewidth,keepaspectratio]{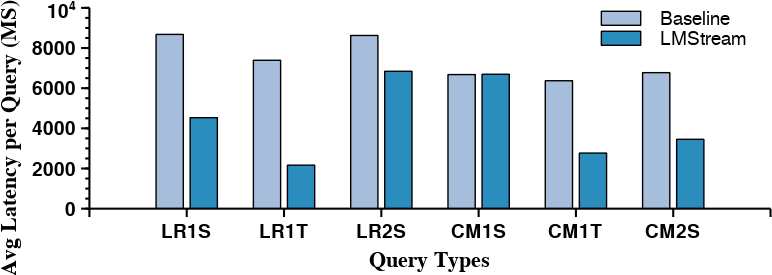}
    \vspace{-0.2in}
    \caption{Average end-to-end latency of all datasets during entire query execution for each workload.}
    \vspace{-0.15in}
\label{fig:avgLatencyQuery}
\end{figure}

\textcolor{black}{
Figure~\ref{fig:avgLatencyQuery} shows the average value of end-to-end latencies of all datasets processed while executing each query. End-to-end latency refers to the time taken from the moment the dataset is entered into the system until the micro-batch execution is completed and goes out to the output stream. The result shows that the average latency of \texttt{LMStream} is much lower in almost all queries compared to \texttt{Baseline}. \texttt{LMStream} shows average latency that is reduced by up to 70.7\% (i.e., in LR1T) compared to the existing method. In the case of \texttt{Baseline}, unconditional buffering causes latency to be consistently above a certain level. Also, the latency continues to increase due to the overall performance degradation caused by buffering (described in Section~\ref{subsec:ourObservations} and Section~\ref{subsec:constructMicroBatch}). In contrast, \texttt{LMStream} dynamically adjusts the buffering phase time according to application type and system performance; therefore, it ensures the bounded latency. When the application performs a sliding window operation on \texttt{LMStream}, it bounds the maximum latency of each micro-batch to the window slide time. Therefore, the average latency value is lower than that. The average latency is much lower when the application performs a tumbling window operation because the system maintains its average value as a minimum. 
}

\begin{figure}
    \centering\includegraphics[width=\linewidth,keepaspectratio]{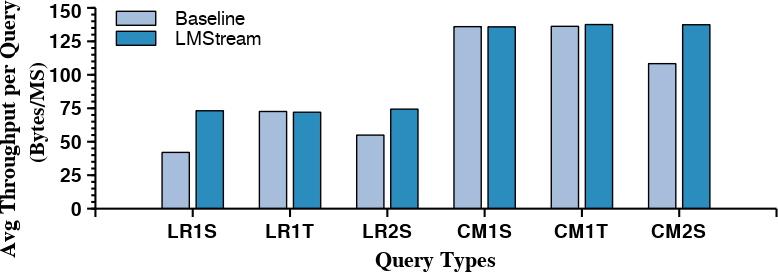}
    \vspace{-0.2in}
    \caption{Average throughput during entire query execution for each workload.}
\label{fig:avgThroughputQuery}
\end{figure}

\textcolor{black}{
Figure~\ref{fig:avgThroughputQuery} shows the average throughput per query in the same experiment. \texttt{LMStream} reduces latency effectively while showing similar or slightly higher throughput than \texttt{Baseline} that uses the throughput-oriented method. It is because efficient query planning can minimize the processing phase time by using dynamic device mapping considering the data size. However, \texttt{Baseline} degrades overall system performance over time due to unconditional buffering. In conclusion, \texttt{LMStream} improves throughput up to 1.74$\times$ (i.e., LR1S) compared to \texttt{Baseline}.
}

\textcolor{black}{
When using a tumbling window like LR1T and CM1T, the latency value of \texttt{Baseline} may appear similar to the trigger time. The reason is that the amount of computation is relatively small, and the trigger time has a dominant influence on the latency. The results indicate that there is no need to use a throughput-oriented method through buffering because the amount of computation is small in the first place. Looking at LR1T and CM1T of Figure~\ref{fig:avgThroughputQuery}, it is shown that the throughput of \texttt{Baseline} and \texttt{LMStream} is bound equally to a specific value. Therefore, \texttt{LMStream}, which maintains both the optimal latency and throughput, is much more efficient. When using a sliding window, the system needs to process larger computation; in this case, unconditional buffering leads to unlimited increase in the latency. Consequently, compared to LR1S, LR2S, and CM2S performed on \texttt{Baseline}, both latency and throughput are further improved when \texttt{LMStream} performs the same query. Although CM1S uses a sliding window operation, the amount of computation was smaller than other queries. In addition, the trigger time of \texttt{Baseline} is the same as the slide time. Thus, \texttt{Baseline} operates with a mechanism similar to \texttt{LMStream} and both shows similar performance results.
}

\subsection{Changes in Maximum Latency and Data Size Over Time}\label{subsec:timelineOfmaximumLatencyAndDataSize}

\begin{figure}[t]
    \centering\includegraphics[width=\linewidth,keepaspectratio]{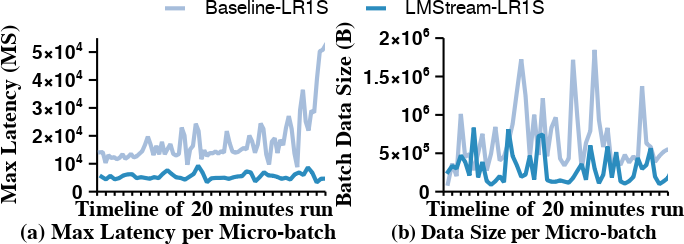}
    \vspace{-0.2in}
    \caption{Timeline during the initial 20-minute run of LR1S: (a) maximum latency per micro-batch, (b) data size per micro-batch.}
    \vspace{-0.2in}
\label{fig:timelineLR1S}
\end{figure}

\begin{figure}[t]
    \centering\includegraphics[width=\linewidth,keepaspectratio]{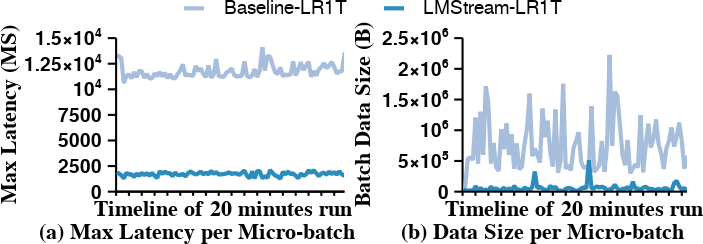}
    \vspace{-0.2in}
    \caption{Timeline during the initial 20-minute run of LR1T: (a) maximum latency per micro-batch, (b) data size per micro-batch.}
\label{fig:timelineLR1T}
\end{figure}

\textcolor{black}{
In this section, we take a closer look at how \texttt{ConstructMicroBatch} works. We selected each workload using the sliding window (LR1S) and the tumbling window (LR1T) and showed the maximum latency per micro-batch over time and the data size processed by the micro-batch. We used random traffic to show more realistic results. We demonstrated the experimental results for the initial 20 minutes run, and then we omitted the afterward results since each system converges in the same pattern.}

\textcolor{black}{
Figure~\ref{fig:timelineLR1S} shows the result when using the sliding window operation. Looking at Figure~\ref{fig:timelineLR1S}(b), since the input data traffic is variable, the size of the data processed for each micro-batch continuously varies in both systems. The difference is that in the case of \texttt{Baseline}, the size of data processed per micro-batch is much larger because it always performs ten seconds of buffering. For this reason, in Figure~\ref{fig:timelineLR1S}(a), the maximum latency per micro-batch of \texttt{Baseline} gradually increases. In contrast, \texttt{LMStream} adjusts the buffering phase time in consideration of the window size and system performance. As a result, maximum latency remains optimal.
}

\textcolor{black}{
Figure~\ref{fig:timelineLR1T} shows the result when using the tumbling window operation. Similar to when using a sliding window, \texttt{LMStream} better controls the buffering phase time to keep maximum latency per micro-batch low. As mentioned in Section~\ref{subsec:overallPerformance}, in the case of a workload using a tumbling window, the amount of computation is relatively small so that even \texttt{Baseline} might escape an unlimited increase in latency. However, if stream processing runs for a long time, the latency will eventually increase.}

\subsection{Effectiveness of Dynamic Device Preference}\label{subsec:effectivenessOfDynamicDevicePreference}

\begin{figure}[t]
    \centering\includegraphics[width=\linewidth,keepaspectratio]{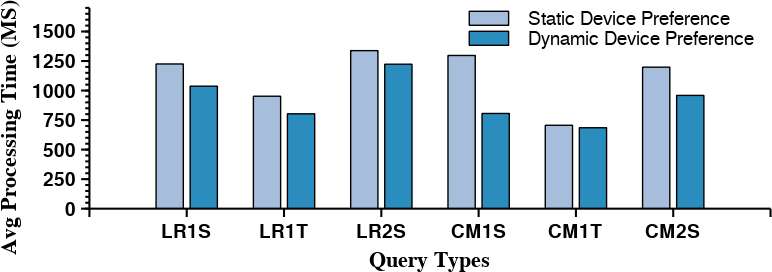}
    \vspace{-0.2in}
    \caption{Average processing phase time per micro-batch during entire query execution for each workload.}
    \vspace{-0.1in}
\label{fig:dynamicDevicePrefEffect}
\end{figure}

\textcolor{black}{
This section presents the efficiency of the query plan reflecting dynamic device preference, which is the core mechanism of \texttt{MapDevice}. The method of reflecting device preference statically was suggested in FineStream~\cite{Fine-Stream}, the most recent study on operation-level query planning in stream processing. To implement a similar method, we experimented with statically fixing the device preference according to Table~\ref{tab:staticDevicePreference} in the \texttt{MapDevice} module. We used random traffic but set the total amount of data to be the same. We averaged the processing phase time required for each micro-batch.
}

\textcolor{black}{
Figure~\ref{fig:dynamicDevicePrefEffect} exhibits the result. The performance is better when \texttt{LMStream} reflects dynamic device preference in all queries. If the size of the data to be processed at once increases, all operations must select the GPU (described in Section~\ref{subsec:mapDevices}). Static device preference does not consider this, and several operations are preferentially placed on the CPU as long as there are available resources. This phenomenon is conspicuous in CM1S. As mentioned in Section~\ref{subsec:overallPerformance}, CM1S operates in a similar way to \texttt{Baseline}. In other words, because data is buffered and processed in this case, the size of data processed in micro-batch increases; therefore, almost all operations must select the GPU. However, when static device preference is enabled, several operations are still executed on the CPU despite increasing data size. Consequently, when performing dynamic device preference during CM1S, performance is improved by up to 37.86\% compared to when static device preference is enabled.
}

\subsection{Overhead Analysis}\label{subsec:overheadAnalysis}

\definecolor{Gray}{gray}{0.9}
\setlength{\extrarowheight}{2pt}
\begin{table}[t]
    \color{black}
    \centering
    \caption{Time ratio required to execute each step: The gray rows are additional overheads in \texttt{LMStream}.}
    \vspace{-0.05in}
    \centering
    \setlength\tabcolsep{2.5pt}
    \begin{tabular}{|c|c|c|c|c|c|c|c|}
        \hline
        \textbf{Ratio} & \textbf{LR1S} & \textbf{LR1T} & \textbf{LR2S} & \textbf{CM1S} & \textbf{CM1T} & \textbf{CM2S} \\ \hline
        \textbf{Buffering Phase} & 18.555 & 28.12 & 93.274 & 51.279 & 13.13 & 77.123 \\ \hline
        \rowcolor{Gray}
        \textbf{Construct Micro-batch} & 0.036 & 0.457 & 0.334 & 0.697 & 0.54 & 0.43 \\ \hline
        \rowcolor{Gray}
        \textbf{Map Device} & 0.182 & 0.014 & 0.003 & 0.006 & 0.0264 & 0.007 \\ \hline
        \textbf{Processing Phase} & 81.011 & 70.45 & 6.387 & 47.925 & 82.648 & 22.437 \\ \hline
        \rowcolor{Gray}
        \textbf{Optimization Blocking} & 0.216 & 0.96 & 0.001 & 0.093 & 3.655 & 0.003 \\ \hline
    \end{tabular}
    \vspace{+0.1in}
\label{tab:overheadBreakdown}
\end{table}

\textcolor{black}{
Finally, we analyzed the overhead caused by the three core mechanisms of \texttt{LMStream}. Table~\ref{tab:overheadBreakdown} is the ratio of time required for each section measured in the experiment to compare overall performance in Section~\ref{subsec:overallPerformance}. Buffering phase time and processing phase time are the times required by default in the micro-batch model, and only the remaining gray-colored rows are additional time spent in \texttt{LMStream}. Overall, the total time ratios occupied by the three mechanisms of \texttt{LMStream} is less than 1\% in most workloads. Therefore, the overhead incurred by \texttt{LMStream} is negligible. For the CM1T, \texttt{LMStream} processes it with minor buffering (i.e., the buffering step of the CM1T takes only about 13.13\% of the time). The shorter the interval at which \texttt{MapDevice} runs, the more often the application will be blocked if regression results have not yet arrived. Nevertheless, the proportion of the total execution time is insignificant at about 3.6\%.}
\section{Conclusion}\label{sec:conclusion}

\textcolor{black}{
In this paper, we present \texttt{LMStream} that proposes dynamic batching and operation-level query planning for GPU-enabled distributed micro-batch stream processing systems. \texttt{LMStream} adjusts the data size of the micro-batch to an appropriate value by considering the window characteristics of the streaming application. As a result, the maximum latency of every micro-batch is bound to the optimal value, effectively reducing the overall latency. In addition, according to the determined micro-batch data size, query operations are dynamically mapped to the appropriate CPU or GPU. Through this, \texttt{LMStream} minimizes the time required for query processing, and as a result, it reduces latency while maximizing throughput. All steps generate minimal overhead and do not interrupt real-time streaming applications. Experimental results show that the mechanisms proposed by \texttt{LMStream} are much more efficient than those of existing methods.
}

%

\Urlmuskip=0mu plus 1mu\relax
\bibliographystyle{bibliography/IEEEtran}
\bibliography{bibliography/References}


\end{document}